\documentclass[10pt,conference]{IEEEtran}
\normalsize

\usepackage{amsmath,amsfonts,amssymb,amsbsy,url,verbatim}
\usepackage{pgfplots}
\usepackage{url}
\usepackage{times}
\usepackage{dsfont}
\usepackage{caption}
\usepackage{subcaption}
\usepackage{cite}
\usepackage{bm}
\usepackage{epstopdf}
\usepackage{algorithm,algorithmic, color}
\usepackage{amsthm}
\usepackage{bm}
\usepackage{graphicx}
\usepackage[T1]{fontenc}

\newcommand{\mat}[1]{\mathbf{#1}}

\newcommand{\bi}{\begin{itemize}}
\newcommand{\ei}{\end{itemize}}
\newcommand{\ben}{\begin{enumerate}}
\newcommand{\een}{\end{enumerate}}
\newcommand{\bc}{\begin{cases}}
\newcommand{\ec}{\end{cases}}
\newcommand{\bd}{\begin{description}}
\newcommand{\ed}{\end{description}}

\newcommand{\be}{\begin{equation}}
\newcommand{\ee}{\end{equation}}
\newcommand{\bea}{\begin{eqnarray}}
\newcommand{\eea}{\end{eqnarray}}

\def \w {\mathbf w}

\def \W {\mathbf W}
\def \I {\mathbf I}

\def \Y {\mathbf Y}
\def \x {\mathbf x}
\def \h {\mathbf h}
\def \y {\mathbf y}
\def \C {\mathbf C}
\def \dbeta {\boldsymbol \beta}
\def \bSigma {\boldsymbol \Sigma}

 \pgfdeclareplotmark{xo}{%
  \pgfusepathqstroke
  \pgfnode{rectangle}{center}{$\star$}{}{}
}

\IEEEoverridecommandlockouts

\begin{document}

\title{CSI-based Outdoor Localization for Massive MIMO: Experiments with a Learning Approach}

\author{Alexis~Decurninge, Luis~Garc\'ia~Ord\'o\~nez,  Paul Ferrand, He Gaoning, Li Bojie, Zhang Wei, Maxime~Guillaud\\
Huawei Technologies}
%email: Contact: \texttt{maxime.guillaud@huawei.com}}

\maketitle

\begin{abstract}
We report on experimental results on the use of a learning-based approach to infer the location of a mobile user of a cellular network within a cell, for a 5G-type Massive multiple input, multiple output (MIMO) system.
We describe how the sample spatial covariance matrix computed from the CSI can be used as the input to a learning algorithm which attempts to relate it to user location.
We discuss several learning approaches, and analyze in depth the application of extreme learning machines, for which theoretical approximate performance benchmarks are available, to the localization problem.
We validate the proposed approach using experimental data collected on a Huawei 5G testbed, provide some performance and robustness benchmarks, and discuss practical issues related to the deployment of such a technique in 5G networks.
\end{abstract}

%\begin{IEEEkeywords}
%\end{IEEEkeywords}

\section{Introduction}

\noindent Future 5G systems are expected to incorporate base stations (BSs) equipped with massive multiple-input multiple-output (MIMO) technology, which uses arrays with the number of antenna elements being some orders of magnitude larger than current state-of-the-art MIMO BSs \cite{Lar14}.
Beyond the expected network-level performance improvements, the presence of excess antennas substantially increases the spatial resolution of the antenna array.
This opens the door to localization methods of increased accuracy based on measuring the propagation characteristics of the wireless channel.
In this article, we investigate to which extent the uplink channel state information (CSI), as estimated by the base station for communications purposes,%---typically, for downlink beamforming---
can be exploited to infer the position of a user equipment (UE)\footnote{Note that we focus on network-side localization here, i.e., the ability for the network to localize a UE, as opposed to user-side localization such as the one enabled by the GPS system, whereby the location information is directly obtained at the UE, but not necessarily shared with the network operator.}.

There is a rich literature on localization strategies based on estimating the state of the wireless propagation channel.
Most of the existing techniques can be understood as follows: the network measures or estimates some propagation parameter---e.g. channel impulse response, propagation delay, angle of arrival or received signal strength---at one or multiple positions in the network, typically BSs or access points (APs).
This \emph{fingerprint} of the user location can be  then either compared to existing databases and maps \cite{Triki_etal_PDP_fingerprinting_VTC2006,Khalajmehrabadi_wlan_fingerprinting_2017}, or subsequently intepreted in a geometric sense.
The assumption of the first approach is that there should be a bijective relationship between the user position and the fingerprint, although this relationship is usually too rich to be modeled and must thus be learned.
A variety of learning algorithms can be applied in this context, among which $K$-nearest neighbors ($K$-nNs), neural network (NNs), support vector machines (SVMs) and Bayesian learning are the most popular.
In contrast, for the geometric approach, the fingerprints are translated into distances and angles from which the user location is triangulated or trilaterated.
In both approaches, the  dimension of the fingerprint is usually the parameter limiting the localization accuracy. 
For example, it was shown in~\cite{Khalajmehrabadi_wlan_fingerprinting_2017} that received signal strength indicator (RSSI)-based approaches become extremely accurate for indoor environments as the number of reference measurements increases.
% However, {\color{red} ([Luis] some drawback should be commented ,e.g., multiple bs vs. single bs)}. {\color{red} ([Luis]  Some comment on the second approach? e.g., geometry-based can be bad???)}.

This limitation can be overcome through the use of multiple antennas at the APs or BSs, as proposed in~\cite{Dayekh_localization_underground_mines_2014,Chapre_CSI_MIMO_2014}.
A similar approach has also been applied in~\cite{Savic_fingerprinting_positioning_distributed_MIMO_2015} to outdoor localization in the so-called \emph{distributed} MIMO case, in which the antenna elements belonging to the same (virtual) BS are not collocated in a single antenna array but distributed within the cell.
%However, multiple antennas have additional benefits for fingerprinting algorithms, as they can provide improved fingerprints with some preprocessing.
%Using the increased spatial resolution coming from the size of the antenna arrays as well as the frequency-domain information obtained through the orthogonal frequency division multiplexing (OFDM) physical layer procedure in modern cellular networks, 
A common way to obtain spatial fingerprints is to extract the cluster of scatterers angle and delay from the raw CSI~\cite{Shutin_angle_delay_2004}.
This fingerprinting method has been studied both in~\cite{Vieira_etal_DCNN_positioning_PIMRC17} and~\cite{Sun_single_site_localization_2018}.
In the former, the authors applied deep neural networks to the obtained fingerprints, while in the latter $K$-nNs type algorithms were considered for their affordable complexity.

In this article, we focus on fingerprinting at a single BS equipped with a large (massive MIMO) antenna array.
We adopt a NN-based learning approach and compare it with a more classical $K$-nN learning algorithm.
In particular, we choose a single-layer random feed-forward neural network, usually referred to as an extreme learning machine (ELM) \cite{huang06,Huang_trends15}.
Beyond its simplicity, the ELM has been proven to exhibit universal approximation capabilities and good generalization performance for a wide range of problems.
In the context of localization, the authors of~\cite{Zou_positioning_procrustes_ELM_TWC16,Zhang_etal_device_free_localization_Sensors17,ELM_positioning_Lu_TransCybernetics17} use RSSI measurements coupled with extreme learning machines for indoor localization.
They however only consider single-antenna elements at the APs. 
The rest of the paper is outlined as follows. We introduce our approach in Section~\ref{section_CSI_based_localization}; the experimental results are presented in Section~\ref{section_experimental}. Finally, in Section~\ref{section_practical_issues} we discuss some general implementation issues that need to be considered before deployment.

\section{Proposed CSI-Based Localization Approach}
\label{section_CSI_based_localization}

\subsection{Input Data Pre-Processing}

\noindent When multiple antennas are used in a single BS at a single cell, some additional pre-processing of the raw channel data is needed to form a useful input for any learning algorithm.
We consider a system where CSI acquisition takes place in the frequency domain, consistent with the OFDM/OFDMA approach being standardized for 5G systems and we let $\h(s,t) \in \mathbb{C}^{N}$ denote the complex vector representing the instantaneous uplink propagation channel between the single-antenna user equipment (UE) and the $N$ BS antennas on subcarrier $s$ at time $t$. Even if the UE position in the horizontal plane (denoted by $\y \in \mathbb{R}^{2}$) is completely static, the instantaneous channel as measured by the baseband processor typically shows some variations due to small-scale fading and to the residual carrier frequency offset resulting from the impossibility of perfectly synchronizing  the oscillators of the transmitter and the receiver.
In order to reduce the noise present in the input data and simultaneously remove the common phase component due to the clock offset, we compute the BS-side channel covariance as
\begin{equation} \label{eq_covariance}
\C (\y) = \mathop{\mathbb{E}}\left[ \h(s,t) \h(s,t)^\dagger \right]
\end{equation}
for some fixed reference subcarrier $s$ and where $(\cdot)^\dagger$ denotes the Hermitian transposition. We implicitly assume that the expectation is computed at time $t$ over a small time horizon $\Delta T$ while the UE remains fixed in a given location $\y$, so that the fading process can be assumed stationary. In practice, the expectation in \eqref{eq_covariance} will be replaced by a sample covariance estimator (see Section~\ref{section_experimental} for more details about our experimental set-up).

In the sequel, we consider as the input to the learning algorithm a vectorized and normalized\footnote{Here we remove the pathloss information (see Section \ref{section_experimental}). Note that there could be alternative approaches to taking the normalized covariance to pre-process the CSI; we did not attempt to exhaustively research this question.} version of the covariance $\C(\y)$, denoted by $ \x \in\mathbb{R}^{N^2}$ and obtained as
\begin{equation} \label{eq:input}
\x = \Big(  \mathrm{vec_{ut}}\big( \mathrm{Re}\big\{\tfrac{\C(\y) }{\text{Tr}(\C(\y))}\big\}\big)^{\mathrm{T}},  \mathrm{vec_{ut}}\big( \mathrm{Im}\big\{\tfrac{\C(\y) }{\text{Tr}(\C(\y))}\big\}\big)^{\mathrm{T}}\Big)^{\mathrm{T}}
\end{equation}
where $\mathrm{vec_{ut}}(\cdot)$ stands for the operator stacking in a vector the upper triangular entries of a $N\times N$ matrix as long as they are non-zero and, hence, removes the inherent redundancy present in $\C(\y)$. This input choice comes from the intuition that the second-order channel statistics capture most of the location-related characteristics of the CSI and, consequently, $\C(\y)$ can be directly related to the UE location $\y$. However, the complexity of the relationship between the spatial scattering environment surrounding a user and the CSI as measured by the BS makes it extremely difficult to express or even model; therefore, we propose to infer it through statistical learning.

\subsection{Learning approaches}

\noindent It is convenient to distinguish between the following two phases of the use of a learning algorithm:
\begin{itemize}
\item[$(\mathrm{i})$] \emph{the training phase}, in which the CSI related input and the UE positions are acquired, i.e., the \emph{training} data set $(\x_j,\y_j)_{j\in\mathcal{T}}$ consisting of $T=|\mathcal{T}|$ pairs $\x_j\in\mathbb{R}^M$, with $M=N^2$, as defined in \eqref{eq:input} and the corresponding user location $\y_j\in\mathbb{R}^2$; this training set is used to infer the mapping relating $\x_j$ to $\y_j$, and
\item[$(\mathrm{ii})$] \emph{the exploitation phase}, in which the mapping obtained in $(\mathrm{i})$ is used to infer the UE position from the observed CSI; the performance of this step is evaluated on a separate \emph{testing} data set  $(\x_j,\y_j)_{j \in \mathcal{T}^{\prime}}$ with $T^{\prime} = |\mathcal{T}^{\prime}|$ data pairs. 
\end{itemize}

The adopted learning approaches are detailed below. 
\medskip
\subsubsection{K-nearest neighbors}
\label{sec:NN}
As a baseline method for comparison purposes, we consider a regression mapping using $K$-nearest neighbors. In this case the training phase only consists in acquiring the training data set. Then, for any observed input data $\x$ during the exploitation phase, we find the $K$ nearest neighbors of $\x$ in the training set, i.e., the $K$ indices $j_1, \ldots, j_K \in \mathcal{T}$  such that, for some distance $d( \cdot ,\cdot )$ defined on the input space (e.g., $d(\x_1,\x_2)=\|\x_1-\x_2\|_2$), we have $d(\x_{j_1},\x) \leq  \cdots \leq  d(\x_{j_K},\x) \leq d(\x_{j},\x)$ for any other $j \in \mathcal{T}$. Finally, the estimated UE location $\hat{\y}$ is computed as the barycenter of the locations of the $K$ nearest neighbors:
\begin{equation}
\hat{\y} = \frac{1}{K} \sum_{k=1}^K\y_{j_k}.
\end{equation}
% Observe that,  instead of giving equal weights $\frac{1}{K}$ to the $K$ closest dictionary elements, one may consider different weights for each one of the $K$ elements; see, e.g., \cite{FDD_covariance_interpolation_Globecom2015} where kernel-based and interpolation-based criteria are studied in a context of covariance transposition.

%We think that this approach in not the best candidate within the context of outdoor single-BS localization due to its lack of robustness to extrapolation, when the position of a user is far from all training data. However, we could not experimentally validate this statement, given the limited set of measurements {\color{red}....???} (see further discussion in Section \ref{section_experimental}).
\smallskip
\subsubsection{Extreme Learning Machine} \label{section_ELM}

% Neural networks compose a typical and powerful other learning approach.
% They are comprised of several neurons connected to each other through a graph (the neurons are therefore the nodes of the graph).
% Each neuron is a nonlinear regressor whose output is a function of all its input values:
% \begin{equation}
% g : \x\in\mathbb{R}^d \mapsto \varphi\Big(\sum_{j=1}^d w_jx_j+w_0\Big)\in\mathbb{R}
% \end{equation}
% where the $w_j$ for $0\leq j\leq d$ are the neuron weights and $\varphi$ is an activation function (see Table~\ref{table:activation} for examples). 
% More specifically, Feedforward Neural Networks (FNNs) are particular neural networks composed by an input, an output and several hidden layers of neurons.
% In typical neural network applications, the first layer correspond to the inputs $\x_j$ and the last layer to the outputs $\y_j$.
% We will consider a specific case of FNN, namely Extreme Learning Machines (ELMs), which use a single-hidden layer (see Fig.~\ref{fig:elm}).

\begin{figure}[t!]
\centering
%\usetikzlibrary{arrows.meta}
\begin{tikzpicture}[
transform shape,
scale=0.8,
y=0.7cm,
>=latex,
every node/.style={circle},
y=0.7cm
]
\def\nI{4}
\def\nC{7}
\def\nO{2}
\pgfmathsetmacro\Ishift{(\nI+1)/2}
\pgfmathsetmacro\Cshift{(\nC+1)/2}
\pgfmathsetmacro\Oshift{(\nO+1)/2}
\foreach \x in {1,...,\nI}{
	
	\draw (-4,\x-\Ishift) node[draw] (I\x) {\phantom{1}};
}
\foreach \x in {1,...,\nC}{
	\draw (0,\x-\Cshift) node[draw] (C\x) {\phantom{1}};
}
\foreach \x in {1,...,\nO}{
	\draw (4,\x-\Oshift) node[draw] (O\x) {\phantom{1}};
}
% Input layers
\foreach \i in {1,...,\nI}{
	\foreach \c in {1,...,\nC}{
	\draw[->] (I\i) -- (C\c);
}
}
\foreach \c in {1,...,\nC}{
	\foreach \o in {1,...,\nO}{
	\draw[->] (C\c) -- (O\o);
}
}
\node at (-2, \Ishift+1) {\large $\mat \W$};
\node at (2, \Ishift+1) {\large $\mat \dbeta$};
%\draw[{Latex[length=2mm]}-{Latex[length=2mm]}] (-5, 0.5-\Ishift) -- (-5, \Ishift-0.5) node[pos=0.5, left] {\Large $p$};
%\draw[{Latex[length=2mm]}-{Latex[length=2mm]}] (5, 0.5-\Oshift) -- (5, \Oshift-0.5) node[pos=0.5, right] {\Large $d$};
\node at (0, -\Cshift) {\large $\varphi(\W^{\mathrm{T}}\x)$};
\node at (-4, -\Cshift) {\large $\x$};
\node at (4, -\Cshift) {\large $\y$};
\end{tikzpicture}
\vspace{-4mm}
\caption{A typical extreme learning machine with $M = 4$ inputs, $n = 8$ neurons and $2$ outputs.}
\label{fig:elm}
\vspace{-2mm}
\end{figure}
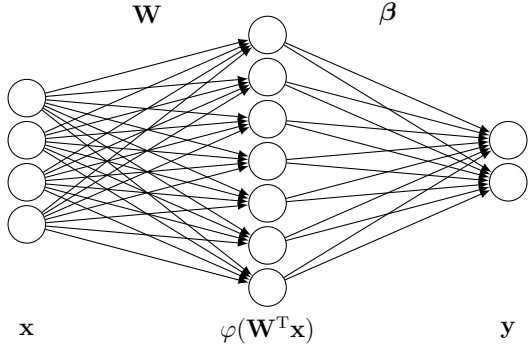

As the main learning method we consider ELMs, which are feedforward neural networks consisting of a random single-hidden layer of neurons (see an example in Fig.~\ref{fig:elm}). An ELM can be understood as a non-linear regression function of the form
\begin{equation}
f :  \x\in\mathbb{R}^{M} \mapsto \sum_{i=1}^n \varphi(\w_i^{\mathrm{T}} \x)\dbeta_i\in\mathbb{R}^2
\end{equation}
where $\w_1,\dots,\w_n\in\mathbb{R}^{M}$ are weights of the $n$ neurons, $\varphi(\cdot)$ is an activation function (see Table~\ref{table:activation} for examples), and  $\dbeta_1,\dots,\dbeta_n\in\mathbb{R}^2$ are the output weights. In ELMs the entries of the weight matrix $\W = (\w_1,\dots\w_n)$ are independently drawn, e.g., from a Gaussian standard distribution and only the output weights $\dbeta=(\dbeta_i)_{i=1\dots n}$ need to be optimized.  

During the training phase, $\dbeta$ is obtained using a simple linear ridge regression between the augmented version of the input, $(\varphi(\w_1^{\mathrm{T}} \x),\dots,\varphi(\w_n^{\mathrm{T}} \x))^{\mathrm{T}} \in \mathbb{R}^n$, and the output $\y$:
\begin{equation}
\label{eq:elm2}
 {\dbeta} = \Y_{\mathcal{T}}(\bSigma_{\mathcal{T}}^{\mathrm{T}}\bSigma_{\mathcal{T}}+\gamma\I_{\mathcal{T}})^{-1}\bSigma_{\mathcal{T}}^{\mathrm{T}}
\end{equation} 
where we denote $\bSigma_{\mathcal{T}}=(\varphi(\w_i^{\mathrm{T}} \x_j))_{i=1,\ldots n, \ j \in \mathcal{T}}\in\mathbb{R}^{n\times T}$ and $\Y_{\mathcal{T}}=(\y_j)_{j \in \mathcal{T}}\in\mathbb{R}^{2\times T}$,
and we have introduced an $L_2$ regularization parameter $\gamma$, which makes the regression more robust without adding complexity.  
%This avoids the slow convergence of classical neural networks using back-propagation algorithms.
In the exploitation phase, the UE location for a new observed input $\x$ is estimated as 
\begin{equation}
\label{eq:elm}
\hat{\y} = \sum_{i=1}^n \varphi(\w_i^{\mathrm{T}} \x){\dbeta}_i={\dbeta}^{\mathrm{T}}\varphi(\W^{\mathrm{T}} \x).
\end{equation}

\begin{table}[b!]
\centering
\begin{tabular}{l|c}
  \hline
  ReLu   & $\varphi(t) = \max(0,t)$\\
  Step   & $\varphi(t) = \mathds{1}_{t\geq 0}$\\
  Sign  & $\varphi(t) = \mathds{1}_{t\geq 0} - \mathds{1}_{t<0}$\\
  \hline
\end{tabular}
\caption{Examples of activation functions.}\label{table:activation}
\end{table}

The choice of ELM as a learning approach is motivated by its theoretical capacity to learn any non-linear mapping, as well as its simplicity \cite{huang06,Huang_trends15} and amenability to theoretical analysis \cite{Louart_etal_RMT_neural_networks_AP17}.

\subsection{Localization Performance}

\noindent In order to analyze the performances of the considered localization approaches, we measure the localization error as
\begin{equation} \label{eq:error}
e(\y_j,\hat{\y}_j) = \|\y_j - \hat{\y}_j\|  
\end{equation}
where $\hat{\y}_j$ is the estimated position for the input $\x_j$. We define the corresponding average training and testing errors as 
%\begin{align}
$E  = \frac{1}{T} \sum_{j \in \mathcal{T}} \|\hat{\y}_j - \y_j\|$ and
$E^{\prime} = \frac{1}{T^{\prime}} \sum_{j \in \mathcal{T}^{\prime}} \|\hat{\y}_j - \y_j\|$.
%\end{align}

In the case of the ELM approach we also introduce the training and testing mean square localization errors:
\begin{align} \label{eq:mse_i}
\mathsf{mse}(\W) & = \frac{1}{T} \sum_{j \in \mathcal{T}} \|\hat{\y}_j - \y_j\|^2 \\
\label{eq:mse_ii}
\mathsf{mse}^{\prime}(\W) & = \frac{1}{T^{\prime}} \sum_{j \in \mathcal{T}^{\prime}} \|\hat{\y}_j - \y_j\|^2 
\end{align}
where we have made explicit their randomness through the dependency on the random weight matrix $\W$.  The authors of \cite{Louart_etal_RMT_neural_networks_AP17} show that for the regression in \eqref{eq:elm}, both training and testing mean square errors almost surely converge to some deterministic limit, i.e.,
\begin{align}
\label{eq:det_eq1}
\mathsf{mse}(\W) - \overline{\mathsf{mse}} &\xrightarrow[ n,M,T \rightarrow \infty]{\mathrm{a.s.}} 0 \\%\qquad \text{and} \qquad 
\mathsf{mse}^{\prime}(\W) - \overline{\mathsf{mse}}^{\prime} &\xrightarrow[ n,M,T^{\prime}\rightarrow \infty]{\mathrm{a.s.}} 0 \label{eq:det_eq2}
\end{align}
where the expressions of the deterministic equivalents $\overline{\mathsf{mse}}$ and $\overline{\mathsf{mse}}^{\prime}$ are given in \cite[Conject.~1 and Thm.~3]{Louart_etal_RMT_neural_networks_AP17}.
Note that $\overline{\mathsf{mse}}$ and $\overline{\mathsf{mse}}^{\prime}$ still depend on the training and testing data sets in order to avoid assuming a generative model of the entries.

The computation of these deterministic values allow us to choose the best hyperparameters, i.e., the regularization parameter $\gamma$, the activation function and the number of neurons independently of the random matrix $\W$.

 % \vspace{-.2cm}

\begin{figure}[t!]
\centering
% {\footnotesize \input{fig_error_map.tikz}}\\
 \includegraphics[width=.9\columnwidth]{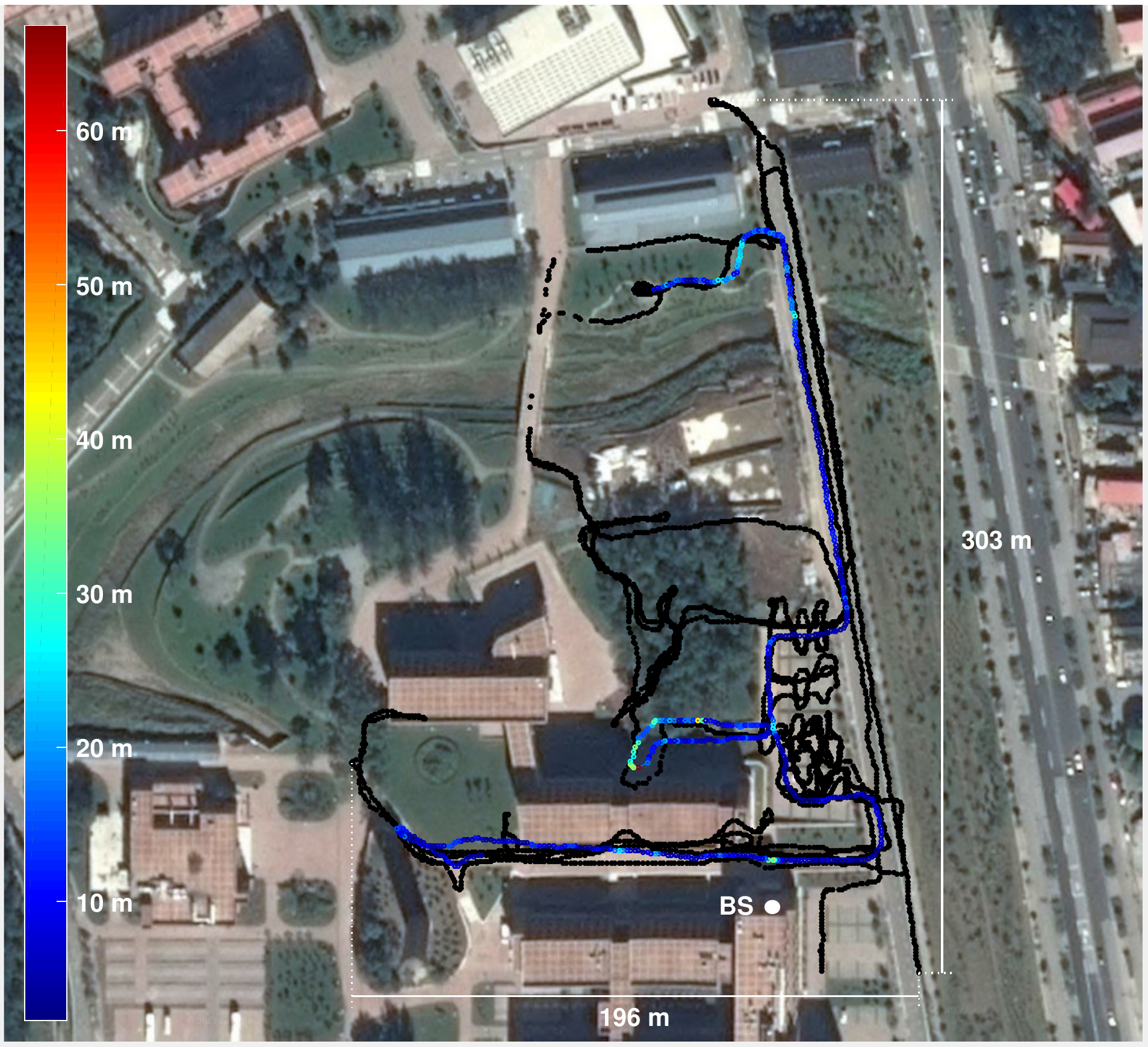}\\
 %\vspace{-1mm}
\caption{Map of the measurement area with the locations of the training set (black dots) and testing data sets (colored dots) with testing location error (using the scale on the left) plotted for the true UE location.
Imagery \textcopyright \ Google, DigitalGlobe.}
\label{fig:error_map}
\vspace{-2mm}
\end{figure}

\section{Experimental Results}
\label{section_experimental}
\noindent Data was collected on the Huawei 5G testbed on the campus of the University of the Chinese Academy of Science in Huairou, P.R. China. CSI data was measured for a single-antenna UE on the uplink of a massive MIMO system using a 32 dual-polarized antenna array at the BS, which is located on the roof of a five-floor building. Due to some malfunctioning RF chains, the measurements from 8 antennas had to be discarded and, hence, unless otherwise stated, $N=56$.   In total, 90 minutes of outdoors measurements were acquired, covering a maximum extent of $196\times303$ meters, depicted on Fig.~\ref{fig:error_map}. Since the mobile UE was moved on a cart, the paths for which data could be recorded was limited to roads and pedestrian walkways. The scenario includes a majority of line-of-sight (LOS) locations, as well as some non-LOS locations. The UE position (latitude/longitude) was recorded using a commercial GPS with 1 Hz sampling rate; it was transformed to a local coordinate system to have dimensions in meters, and subsequently linearly interpolated in time when required. The CSI sampling rate was 10 Hz. Path-loss information was not deemed reliable due to automatic gain control, and therefore all covariance matrices were normalized as in \eqref{eq:input}. Gaps in the trace are due to the UE being out of range from the BS.

Clearly, when the UE is moving, $\h(s,t)$ can not be considered to be stationary and the statistics cannot be related to the position. This makes it difficult to evaluate the expectation when computing in practice the covariance as defined in \eqref{eq_covariance}. In order to escape this apparent contradiction and to be able to process the experimental data acquired for a moving user, we apply a sample covariance estimator over a 1-second horizon,\footnote{Note that at pedestrian speeds, $1$ s translates into approximately $1$ m, which is small with respect to the size of the considered area, and roughly commensurate with the expected accuracy of the GPS location data.} where the statistics can be assumed approximately constant. In the experiments presented here, a single subcarrier (always the same) is used, since including multiple ones have shown little or no gain, even when the selected subcarriers were spaced apart within the available 10 MHz band. We speculate that this is due to the lack of channel frequency selectivity in the considered scenario. 

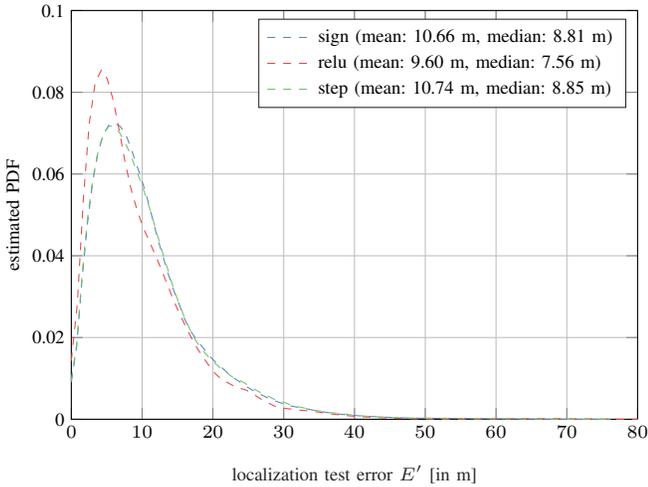
\begin{figure} [t!]
\centering
{\scriptsize % This file was created by matlab2tikz.
\definecolor{mycolor1}{rgb}{0.34667,0.53600,0.69067}%
\definecolor{mycolor2}{rgb}{0.91529,0.28157,0.28784}%
\definecolor{mycolor3}{rgb}{0.44157,0.74902,0.43216}%
\begin{tikzpicture}

\begin{axis}[%
width=0.85\columnwidth,
height=0.614\columnwidth,
at={(0\textwidth,0\textwidth)},
scale only axis,
xmin=0,
xmax=80,
xlabel style={font=\color{white!15!black}},
xlabel={localization test error $E^{\prime}$ [in m]},
ymin=0,
ymax=0.1,
ylabel style={font=\color{white!15!black}},
ylabel={estimated PDF},
axis background/.style={fill=white},
xmajorgrids,
ymajorgrids,
ylabel near ticks,
ytick={0,0.02,0.04,0.06,0.08,0.1},
yticklabels={{0},{0.02},{0.04},{0.06},{0.08},{0.1}},
legend style={legend cell align=left, align=left, draw=white!15!black}
]
\addplot [color=mycolor1, dashed]
  table[row sep=crcr]{%
0	0.00907799369115558\\
0.762	0.0178417491391558\\
1.524	0.0343832020997375\\
2.286	0.0480074165041296\\
3.048	0.0583688506826555\\
3.81	0.0656757928194756\\
4.572	0.0698198848997087\\
5.334	0.0720520600062607\\
6.096	0.0724264971465722\\
6.858	0.0717342098292759\\
7.62	0.069998073635291\\
8.382	0.0665402489826386\\
9.144	0.0626754797852103\\
9.906	0.058780611139204\\
10.668	0.054365624021768\\
11.43	0.0493053047268174\\
12.192	0.0442473933877531\\
12.954	0.0397553516819572\\
13.716	0.0354246430205399\\
14.478	0.0314635555876617\\
15.24	0.0281080690601748\\
16.002	0.0245310505911532\\
16.764	0.0218004286161477\\
17.526	0.0198367405909124\\
18.288	0.0181692311396855\\
19.05	0.0165799802547617\\
19.812	0.0149293264947386\\
20.574	0.0136783934118327\\
21.336	0.0125081268511161\\
22.098	0.0113282284668545\\
22.86	0.0102193647812372\\
23.622	0.00934888872835849\\
24.384	0.00844470129307231\\
25.146	0.00754894170338799\\
25.908	0.00677598786390233\\
26.67	0.00612343181872908\\
27.432	0.0054889354427027\\
28.194	0.00481109586072384\\
28.956	0.00438729562474415\\
29.718	0.003929784006357\\
30.48	0.00341809338052927\\
31.242	0.00317609381396133\\
32.004	0.00291723855618966\\
32.766	0.0026595872763611\\
33.528	0.00246333887163187\\
34.29	0.00217197620939584\\
35.052	0.0019299766428279\\
35.814	0.0017301163042693\\
36.576	0.00152423607599509\\
37.338	0.00140745021551205\\
38.1	0.00121842567844157\\
38.862	0.00107154036938043\\
39.624	0.000994485781020491\\
40.386	0.000882515832309952\\
41.148	0.00077295383948566\\
41.91	0.000700715162898219\\
42.672	0.000659779912832\\
43.434	0.000573093500927063\\
44.196	0.000526138361145224\\
44.958	0.000474367309590888\\
45.72	0.00041898432420718\\
46.482	0.000390088853572202\\
47.244	0.000344337691733487\\
48.006	0.000307014375496641\\
48.768	0.000290158684292904\\
49.53	0.000267283103373546\\
50.292	0.00023597967685232\\
51.054	0.000238387632738568\\
51.816	0.000238387632738568\\
52.578	0.000207084206217342\\
53.34	0.000195044426786101\\
54.102	0.000190228515013605\\
54.864	0.000185412603241109\\
55.626	0.000179392713525489\\
56.388	0.000168556912037371\\
57.15	0.000146885309061138\\
57.912	0.000132437573743649\\
58.674	0.000138457463459269\\
59.436	0.000125213706084904\\
60.198	0.00011196994871054\\
60.96	9.39102795636783e-05\\
61.722	9.99301692792987e-05\\
62.484	9.02983457343061e-05\\
63.246	5.89949192130803e-05\\
64.008	6.01988971562044e-05\\
64.77	4.93630956680873e-05\\
65.532	3.85272941799706e-05\\
66.294	3.0099448578102e-05\\
67.056	1.80596691468612e-05\\
67.818	1.32437573743649e-05\\
68.58	1.56517132606131e-05\\
69.342	8.42784560186857e-06\\
70.104	3.61193382937224e-06\\
70.866	6.01988971562041e-06\\
71.628	3.61193382937224e-06\\
72.39	1.20397794312408e-06\\
73.152	3.61193382937224e-06\\
73.914	2.40795588624816e-06\\
74.676	0\\
75.438	1.20397794312408e-06\\
76.2	2.40795588624816e-06\\
};
\addlegendentry{sign (mean: 10.66 m, median: 8.81 m)}

\addplot [color=mycolor2, dashed]
  table[row sep=crcr]{%
0	0.0143358123880354\\
0.845	0.0281515661473319\\
1.69	0.0537169534770099\\
2.535	0.0724629498941426\\
3.38	0.0824548070137343\\
4.225	0.0853699581998806\\
5.07	0.0832875522501493\\
5.915	0.0780891373975354\\
6.76	0.0707714022040063\\
7.605	0.0627848650996146\\
8.45	0.0562173606210303\\
9.295	0.051103631724662\\
10.14	0.0469464198469139\\
10.985	0.0436349818142338\\
11.83	0.0403398295423701\\
12.675	0.0367971337060963\\
13.52	0.0331545518701482\\
14.365	0.0294240269257912\\
15.21	0.0260593887411107\\
16.055	0.023086694533413\\
16.9	0.0201302860865317\\
17.745	0.0174800499429998\\
18.59	0.0152098148851854\\
19.435	0.0131730090657402\\
20.28	0.0112599750285001\\
21.125	0.00990934259812172\\
21.97	0.00906356875305359\\
22.815	0.00825470929916942\\
23.66	0.00768687910536887\\
24.505	0.00725584930242659\\
25.35	0.00659247597850279\\
26.195	0.00558493024265783\\
27.04	0.00467401335432388\\
27.885	0.00390858259595027\\
28.73	0.00321589490255686\\
29.575	0.00281092231692091\\
30.42	0.00260463601324575\\
31.265	0.00246023560067314\\
32.11	0.00227457792736551\\
32.955	0.00212257749307855\\
33.8	0.00199989142826123\\
34.645	0.0018196623418924\\
35.49	0.00162314749470712\\
36.335	0.00140600401715434\\
37.18	0.00119428912654036\\
38.025	0.00105097443135552\\
38.87	0.00090331686661962\\
39.715	0.000766516475761357\\
40.56	0.000666630476087072\\
41.405	0.000559144454698443\\
42.25	0.000498344280983661\\
43.095	0.000453829868085338\\
43.94	0.000374572498778568\\
44.785	0.000293143694696271\\
45.63	0.000226914934042668\\
46.475	0.000201943434124097\\
47.32	0.000172629064654471\\
48.165	0.000122686064817328\\
49.01	9.12002605721732e-05\\
49.855	6.18858911025461e-05\\
50.7	3.36572390206829e-05\\
51.545	1.84571955919873e-05\\
52.39	1.41143260409316e-05\\
53.235	1.19428912654036e-05\\
54.08	1.41143260409316e-05\\
54.925	1.84571955919874e-05\\
55.77	1.95429129797514e-05\\
56.615	2.38857825308072e-05\\
57.46	2.38857825308072e-05\\
58.305	3.3657239020683e-05\\
59.15	5.10287172249064e-05\\
59.995	5.64573041637263e-05\\
60.84	8.03430866945335e-05\\
61.685	0.000103143151837576\\
62.53	0.00010965745616416\\
63.375	0.00011508604310298\\
64.22	0.00011508604310298\\
65.065	0.000135714673470496\\
65.91	0.000137886108246024\\
66.755	0.000119428912654036\\
67.6	0.00012594321698062\\
68.445	0.000138971825633788\\
69.29	0.000128114651756148\\
70.135	0.000100971717062048\\
70.98	8.14288040822967e-05\\
71.825	8.46859562455894e-05\\
72.67	8.90288257966453e-05\\
73.515	6.5143043265838e-05\\
74.36	4.77715650616145e-05\\
75.205	4.12572607350307e-05\\
76.05	2.93143694696271e-05\\
76.895	2.17143477552793e-05\\
77.74	1.84571955919874e-05\\
78.585	1.52000434286955e-05\\
79.43	9.7714564898757e-06\\
80.275	5.4285869388198e-06\\
81.12	6.51430432658376e-06\\
81.965	4.34286955105587e-06\\
82.81	0\\
83.655	1.08571738776397e-06\\
84.5	2.17143477552793e-06\\
};
\addlegendentry{relu (mean: 9.60 m, median: 7.56 m)}

\addplot [color=mycolor3, dashed]
  table[row sep=crcr]{%
0	0.00951365239416819\\
0.757	0.0184164919467235\\
1.514	0.0353011040684498\\
2.271	0.0489850084229152\\
3.028	0.0587677093306509\\
3.785	0.0656599566128974\\
4.542	0.0698714142013986\\
5.299	0.0716893095633415\\
6.056	0.0717426344939586\\
6.813	0.0700895616448317\\
7.57	0.0677832583956467\\
8.327	0.0649012882818465\\
9.084	0.0613527565353339\\
9.841	0.0579266297431919\\
10.598	0.0540969301806988\\
11.355	0.0499654599881231\\
12.112	0.0453140717220317\\
12.869	0.0408759831784083\\
13.626	0.0366621017294244\\
14.383	0.0323342988377589\\
15.14	0.028677905299771\\
15.897	0.0253026795777635\\
16.654	0.0223492055797268\\
17.411	0.0197181050258747\\
18.168	0.0177390229418395\\
18.925	0.0161271557209167\\
19.682	0.0146279980124344\\
20.439	0.0134851477948929\\
21.196	0.0123604765309709\\
21.953	0.0113824488262456\\
22.71	0.0103983614703138\\
23.467	0.00973179983760135\\
24.224	0.00912825857743627\\
24.981	0.00830293408311418\\
25.738	0.00745943063517264\\
26.495	0.00672984862991287\\
27.252	0.00610328069516319\\
28.009	0.00545247415558759\\
28.766	0.00481136305794239\\
29.523	0.00432053131021779\\
30.28	0.00400906523820489\\
31.037	0.00349399488565438\\
31.794	0.00314495897616133\\
32.551	0.00290378485814357\\
33.308	0.00254262964623757\\
34.065	0.0022481305976028\\
34.822	0.00210269896864736\\
35.579	0.00193060487438344\\
36.336	0.00165549670960939\\
37.093	0.00155854229030577\\
37.85	0.00146279980124344\\
38.607	0.00130161307915117\\
39.364	0.00110043265909614\\
40.121	0.000959848751105884\\
40.878	0.000876225564456512\\
41.635	0.000758668331050863\\
42.392	0.000665349702471124\\
43.149	0.000599905469441181\\
43.906	0.000572031073891386\\
44.663	0.0004908317477246\\
45.42	0.000429023305418541\\
46.177	0.000420539793729474\\
46.934	0.000380546095766727\\
47.691	0.000332068886114914\\
48.448	0.000302982560323829\\
49.205	0.000307830281289008\\
49.962	0.000302982560323827\\
50.719	0.000294499048634761\\
51.476	0.00026904851356756\\
52.233	0.000271472374050149\\
52.99	0.000262988862361082\\
53.747	0.000237538327293882\\
54.504	0.000239962187776471\\
55.261	0.000213299722467974\\
56.018	0.000196332699089841\\
56.775	0.000191484978124659\\
57.532	0.00015755093136839\\
58.289	0.000149067419679324\\
59.046	0.000151491280161914\\
59.803	0.00013694811726637\\
60.56	0.000117557233405645\\
61.317	0.000101802140268806\\
62.074	9.69544193036247e-05\\
62.831	7.02919539951278e-05\\
63.588	6.05965120647657e-05\\
64.345	5.69607213408796e-05\\
65.102	4.2417558445336e-05\\
65.859	3.15101862736782e-05\\
66.616	2.66624653084967e-05\\
67.373	2.18147443433156e-05\\
68.13	1.21193024129532e-05\\
68.887	9.69544193036242e-06\\
69.644	8.48351168906719e-06\\
70.401	3.63579072388598e-06\\
71.158	2.42386048259061e-06\\
71.915	4.84772096518126e-06\\
72.672	3.63579072388595e-06\\
73.429	2.42386048259061e-06\\
74.186	1.2119302412953e-06\\
74.943	1.2119302412953e-06\\
75.7	2.42386048259061e-06\\
};
\addlegendentry{step (mean: 10.74 m, median: 8.85 m)}

\end{axis}
\end{tikzpicture}%
}
\vspace{-1mm}
\caption{Localization error distribution. } \label{fig:error_distribution}
\end{figure}

\begin{figure} [t!]
\vspace{-2mm}
\centering
{\scriptsize % This file was created by matlab2tikz.
\definecolor{mycolor1}{rgb}{0.34667,0.53600,0.69067}%
\definecolor{mycolor2}{rgb}{0.91529,0.28157,0.28784}%
\definecolor{mycolor3}{rgb}{0.44157,0.74902,0.43216}%
\begin{tikzpicture}

\begin{axis}[%
width=0.85\columnwidth,
height=0.614\columnwidth,
at={(0\columnwidth,0\columnwidth)},
scale only axis,
xmin=8,
xmax=56,
xtick={4,8,12,16,20,24,28,32,36,40,44,48,52,56},
xlabel style={font=\color{white!15!black}},
xlabel={number of antennas $N$},
ymin=5.5,
ymax=20.5,
ytick={6,8,10,12,14,16,18,20},
ylabel style={font=\color{white!15!black}},
ylabel={localization errors $E$ and $E^{\prime}$ [in m]},
axis background/.style={fill=white},
ylabel near ticks,
xlabel near ticks,
xmajorgrids,
ymajorgrids,
legend style={legend cell align=left, align=left, draw=white!15!black}
]
\addplot [color=mycolor1]
  table[row sep=crcr]{%
4	22.4271278477446\\
8	16.1534784929002\\
12	10.6451492905843\\
16	9.48918154886481\\
20	8.96648362325765\\
24	8.53977431600507\\
28	8.24334016368682\\
32	8.06027340363413\\
36	7.82512720587548\\
40	7.66827606784705\\
44	7.61265717282395\\
48	7.54892326541884\\
52	7.51511873083816\\
56	7.48538427780745\\
};
\addlegendentry{sign (train)}

\addplot [color=mycolor1, dashed]
  table[row sep=crcr]{%
4	21.355443800048\\
8	20.6640942902279\\
12	13.4801914829367\\
16	13.6223033258548\\
20	13.3641271531065\\
24	12.605137879474\\
28	11.6435486559552\\
32	11.3059527832098\\
36	11.1534594859577\\
40	11.1079473220823\\
44	10.8636353444062\\
48	10.7922340737833\\
52	10.8067718555684\\
56	10.6573054446781\\
};
\addlegendentry{sign (test)}

\addplot [color=mycolor2]
  table[row sep=crcr]{%
4	25.0805009661013\\
8	15.9501453564457\\
12	10.0351506209907\\
16	8.71846292744077\\
20	7.96131905706027\\
24	7.38576330406723\\
28	7.01780780739275\\
32	6.75457996877292\\
36	6.46059202034856\\
40	6.20635117658689\\
44	6.06333156958398\\
48	5.92713095587611\\
52	5.8260364492165\\
56	5.74820059234031\\
};
\addlegendentry{relu (train)}

\addplot [color=mycolor2, dashed]
  table[row sep=crcr]{%
4	21.5767658764411\\
8	18.5326930646027\\
12	13.2263114395537\\
16	13.5869621828886\\
20	12.9110936331156\\
24	12.1584152774544\\
28	11.3627865510896\\
32	11.1670291865853\\
36	10.8872593527505\\
40	10.3814045871137\\
44	10.030126025334\\
48	9.93436089397673\\
52	9.77617647727919\\
56	9.5799627262209\\
};
\addlegendentry{relu (test)}

\addplot [color=mycolor3]
  table[row sep=crcr]{%
4	22.182290636758\\
8	15.8883401025012\\
12	10.4950170501906\\
16	9.33210730849217\\
20	8.78692024258277\\
24	8.33365214454234\\
28	8.02989831129014\\
32	7.83974330194024\\
36	7.60573815448679\\
40	7.44694788033734\\
44	7.38660783296798\\
48	7.30745023668507\\
52	7.27104016014948\\
56	7.23858590141182\\
};
\addlegendentry{step (train)}

\addplot [color=mycolor3, dashed]
  table[row sep=crcr]{%
4	21.3067243085343\\
8	20.3862953531528\\
12	13.511357428922\\
16	13.706447422511\\
20	13.5019502780698\\
24	12.7044234285295\\
28	11.7000057713633\\
32	11.3455315430446\\
36	11.187043869239\\
40	11.1393943193302\\
44	10.8980752975918\\
48	10.8409376593521\\
52	10.8718880891082\\
56	10.7406551165099\\
};
\addlegendentry{step (test)}

\addplot [color=mycolor1, draw=none, mark=o, mark options={solid, mycolor1}, forget plot]
  table[row sep=crcr]{%
4	22.4271278477446\\
8	16.1534784929002\\
12	10.6451492905843\\
16	9.48918154886481\\
20	8.96648362325765\\
24	8.53977431600507\\
28	8.24334016368682\\
32	8.06027340363413\\
36	7.82512720587548\\
40	7.66827606784705\\
44	7.61265717282395\\
48	7.54892326541884\\
52	7.51511873083816\\
56	7.48538427780745\\
};
\addplot [color=mycolor1, draw=none, mark=o, mark options={solid, mycolor1}, forget plot]
  table[row sep=crcr]{%
4	21.355443800048\\
8	20.6640942902279\\
12	13.4801914829367\\
16	13.6223033258548\\
20	13.3641271531065\\
24	12.605137879474\\
28	11.6435486559552\\
32	11.3059527832098\\
36	11.1534594859577\\
40	11.1079473220823\\
44	10.8636353444062\\
48	10.7922340737833\\
52	10.8067718555684\\
56	10.6573054446781\\
};
\addplot [color=mycolor2, draw=none, mark=o, mark options={solid, mycolor2}, forget plot]
  table[row sep=crcr]{%
4	25.0805009661013\\
8	15.9501453564457\\
12	10.0351506209907\\
16	8.71846292744077\\
20	7.96131905706027\\
24	7.38576330406723\\
28	7.01780780739275\\
32	6.75457996877292\\
36	6.46059202034856\\
40	6.20635117658689\\
44	6.06333156958398\\
48	5.92713095587611\\
52	5.8260364492165\\
56	5.74820059234031\\
};
\addplot [color=mycolor2, draw=none, mark=o, mark options={solid, mycolor2}, forget plot]
  table[row sep=crcr]{%
4	21.5767658764411\\
8	18.5326930646027\\
12	13.2263114395537\\
16	13.5869621828886\\
20	12.9110936331156\\
24	12.1584152774544\\
28	11.3627865510896\\
32	11.1670291865853\\
36	10.8872593527505\\
40	10.3814045871137\\
44	10.030126025334\\
48	9.93436089397673\\
52	9.77617647727919\\
56	9.5799627262209\\
};
\addplot [color=mycolor3, draw=none, mark=o, mark options={solid, mycolor3}, forget plot]
  table[row sep=crcr]{%
4	22.182290636758\\
8	15.8883401025012\\
12	10.4950170501906\\
16	9.33210730849217\\
20	8.78692024258277\\
24	8.33365214454234\\
28	8.02989831129014\\
32	7.83974330194024\\
36	7.60573815448679\\
40	7.44694788033734\\
44	7.38660783296798\\
48	7.30745023668507\\
52	7.27104016014948\\
56	7.23858590141182\\
};
\addplot [color=mycolor3, draw=none, mark=o, mark options={solid, mycolor3}, forget plot]
  table[row sep=crcr]{%
4	21.3067243085343\\
8	20.3862953531528\\
12	13.511357428922\\
16	13.706447422511\\
20	13.5019502780698\\
24	12.7044234285295\\
28	11.7000057713633\\
32	11.3455315430446\\
36	11.187043869239\\
40	11.1393943193302\\
44	10.8980752975918\\
48	10.8409376593521\\
52	10.8718880891082\\
56	10.7406551165099\\
};
\end{axis}
\end{tikzpicture}%
}
\vspace{-1mm}
\caption{Average localization error vs. number of antennas}  \label{fig:error_vs_Nantennas}
\vspace{-2mm}
\end{figure}
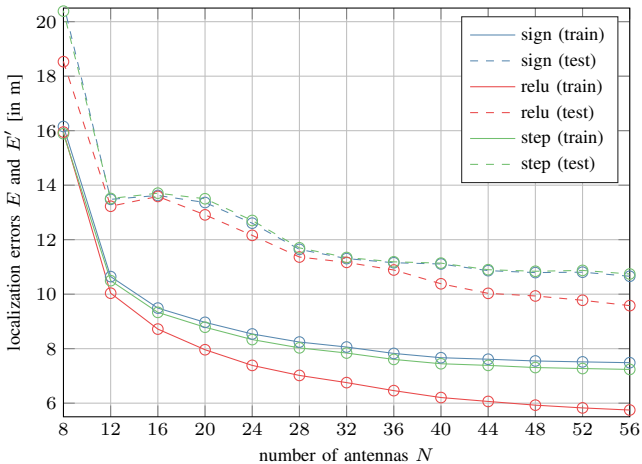

In Fig.~\ref{fig:error_map} we present the resulting localization error as defined in \eqref{eq:error} for the ELM approach with $n=10^4$, a ReLu activation function, and $\W$ with i.i.d.~standard normal entries. The black dots depict the training data set of size $T=3500$, while the colored dots show the localization error for each location within the testing set of size $T^{\prime}=550$, using the scale on the left. Observe that the accuracy lies mostly below 10 m for most locations, while some loss of accuracy is noticeable for locations far away from the BS (where the CSI estimates are more noisy due to a lower SNR) and for a set of NLOS locations where the BS is shadowed by a tall (six floors) building. This can be further confirmed by the testing error distribution estimated over 1000 realizations of $\W$ in Fig.~\ref{fig:error_distribution}.

\subsection{Influence of the number of antennas}

\noindent In order to evaluate the sensitivity of the localization error to the number of antennas, we ran simulations with varying number of BS antennas by discarding part of the data. For each value of $N$, the subset of antenna elements in use was chosen so as to heuristically maximize the inter-element spacing, and the polarization diversity. The result is shown in Fig.~\ref{fig:error_vs_Nantennas}. Our experiments indicate that the localization error is low and exhibits little sensitivity to $N$ as long as $N\geq 12$, while for $N \leq 8$ the localization accuracy decreases significantly. This indicates that the potential applications of the proposed method are not limited to the massive MIMO regime, but also extends to the typical number of antennas found in 4.5G MIMO systems.

\subsection{Optimization of the ELM hyperparameters}
\begin{figure*}[t!] 
\begin{subfigure}[t!]{0.47\textwidth} 
                {\scriptsize \input{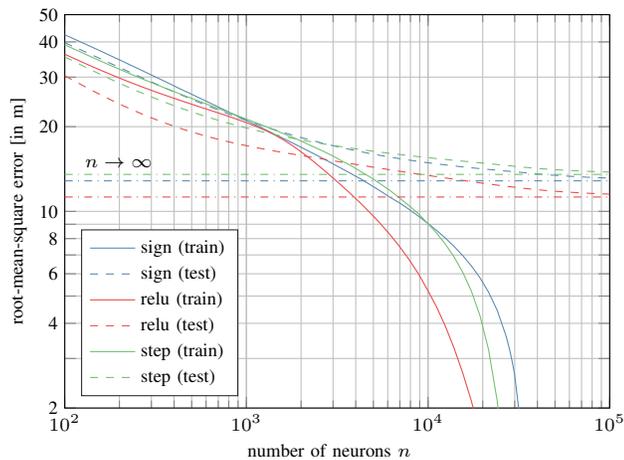}}
								\subcaption{$\sqrt{\overline{\mathsf{mse}}}$ (solid) and $\sqrt{\overline{\mathsf{mse}}^{\prime}}$ (dashed) vs number of neurons.}\label{fig_mse_vs_neurons}
        \end{subfigure} \hspace{0.5cm}        
        \begin{subfigure}[t!]{0.47\textwidth} 
               {\scriptsize \input{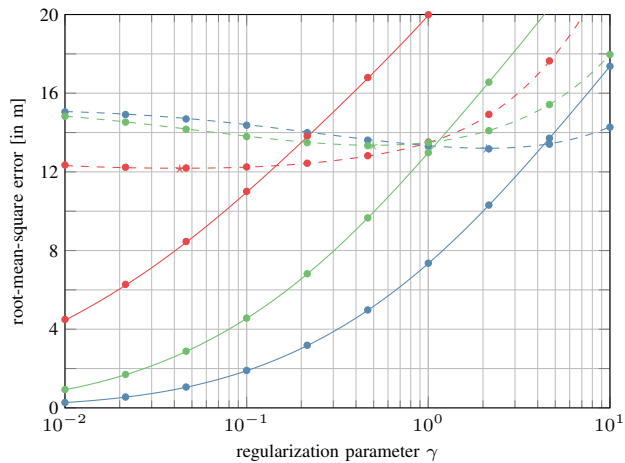}}
							\subcaption{$\sqrt{\overline{\mathsf{mse}}}$ (solid) and $\sqrt{\overline{\mathsf{mse}}^{\prime}}$ (dashed) vs. regularization parameter.} \label{fig_mse_vs_gamma}
        \end{subfigure}
        \caption{Optimization of the neural network parameters based on the deterministic equivalent analysis of the ELM.} \label{fig:error_rmt}
 \vspace{-2mm}
\end{figure*}

\noindent The theoretical framework of \cite{Louart_etal_RMT_neural_networks_AP17} was used to optimize certain parameters of the ELM (number of neurons $n$, activation function, and regularization parameter $\gamma$) based on the training and testing data sets. The results are presented in Fig.~\ref{fig:error_rmt}, where it can be noticed that the ReLu activation function uniformly outperforms the sign and step functions. 

The dependency on the number of neurons $n$ is shown in Fig.~\ref{fig_mse_vs_neurons}. Both deterministic equivalents for the training and testing MSEs decrease when $n$ grows. The sharp drop in the training MSE that can be observed for $n\geq 10^4$ corresponds to $n\geq T$. Note, however, that no adverse over-fitting effect is observed for large $n$ (the testing MSEs keep monotonically decreasing towards the asymptotic values of $\overline{\mathsf{mse}}^{\prime}$ plotted in dash-dotted lines). The curves of Fig.~\ref{fig_mse_vs_neurons} have been generated by computing the optimum $\gamma$ minimizing $\overline{\mathsf{mse}}^{\prime}$ for each $n$.

Fig.~\ref{fig_mse_vs_gamma} shows the influence of the regularization parameter $\gamma$ on the localization MSE, for the three considered activation functions. Here, we fixed $n=10^4$ since it provides a good tradeoff between complexity and performance. The dots correspond to the experimental values of $\mathbb{E}_{\W}\{\mathsf{mse}(\W)\}$ and  $\mathbb{E}_{\W}\{\mathsf{mse}^{\prime}(\W)\}$  directly computed from running the trained ELM on the training and testing data for different realizations of $\W$, while the solid and dashed lines show the corresponding deterministic equivalents (see~\eqref{eq:det_eq1} and \eqref{eq:det_eq2}). It can be seen that there is an excellent match between the deterministic equivalents and the experimental values.

\subsection{Comparison of ELM with $K$-nN approach}
\noindent  In order to benchmark our results, we compare in Table~\ref{table:NNELM} the ELM learning approach with the $K$-nN approach presented in Section~\ref{sec:NN} wherein the chosen distance in the input space is the $2$-norm. As we see in Table~\ref{table:NNELM}, $K$-nN outperforms the ELM-based approach in terms of the average accuracy, while at the same time being more prone to catastrophic errors. However, at this stage we refrain from drawing conclusions from this comparison, since the set-up is unfavorable to the ELM approach due to the small size of the training dataset (with respect to the dimension of the data). Furthermore, the dataset is very structured in the sense  that it consists only of pathways, and this guarantees that for every data point in the testing data set there are always a couple of ``good'' nearest neighbors, which results in the best situation for the $K$-nN algorithm. Finally, we would like also to remark that the computational complexity for the $K$-nN localization strategy grows with the size of the training data set, whereas in the ELM case is only determined by the number of neurons. This can render the $K$-nN approach unfeasible is practice.

\begin{table}[b!]
\centering
\begin{tabular}{l |c c}
  \hline
Statistic & ELM (ReLu) & 3-nN\\
  \hline
  Average localization error [in m] & 9.60 &  8.16\\
  Median localization error [in m] & 7.56 & 4.29\\
  Maximum localization error [in m] & 73.26 & 108.25\\
  \hline
\end{tabular}
\caption{Comparison of statistics computed on the testing set on localization error between $3$-nNs and ELM.} \label{table:NNELM}
\end{table}

\section{Practical implementation issues}
\label{section_practical_issues}
\noindent  Here we discuss the practical issues related to the implementation of a learning-based localization approach in a cellular network. In particular, the availability of training data, and potentially the necessity to periodically update this training, are important questions in the context of the proposed application.

\subsection{Potential training approaches}
\noindent The method outlined in Section~\ref{section_CSI_based_localization} relies on the availability of a training dataset for which CSI has been obtained for a large number of reference positions, i.e., we need to assume that a reference position information is available for this data set. 

Since the mapping between CSI and location is strongly dependent on the geometric properties of the scattering environment which surrounds the BS (i.e. shape and location of buildings and other macroscopic obstacles), it is expected that a dedicated training must be performed for each BS site. This also indicates that additional training might be required if significant changes occur in the scattering environment (new buildings...).
Therefore, it is important to identify practically feasible strategies to deal with the acquisition (and potentially, with the updates) of the training data set. Among the possible approaches, we distinguish between operator-based  and crowdsourcing-based training.

\paragraph{Operator-based training}
In this strategy, the training data is acquired through dedicated measurements performed by the operator. The location of the reference UE is determined either through GPS, or through other kinds of geometric measures (distance to a reference point in the horizontal plane, elevation based on the floor number in case of 3-D localization in indoor environments...); this operation can be performed by a human or robotic operator.
Although it is clearly a costly approach, it comes with the guarantee that the training data set is accurate and exhaustive.

\paragraph{Crowdsourcing-based training}
In this approach, the location data is collected by a subset of the end users of the cellular network. Practically, this can take the form of an operator-provided application which runs on the users' mobile device and relies on its existing localization service (GPS-based or other) to determine its position, and feed it back to the operator; the operator can provide a reward (monetary, or through a discount or additional services) in exchange for the voluntary participation of the user to this crowdsourcing data collection. For a given user participating in the crowdsourcing, this data collection does not have to be permanent, so that the associated impact on power and data consumption is minimized.
In that scenario, the operator has no control over the geographical area covered by the training data, since it merely follows the end user's behavior; thus, it can not be ensured that the training data covers uniformly and/or exhaustively the geographical area covered by the cell. On the other hand, it can be argued that, if sufficiently many users can be enrolled in the crowdsourcing data collection, the distribution of user locations in the training data will match the real user distribution, ensuring a faithful location estimate during the exploitation phase.

\subsection{Training Update}
\noindent In the context of the considered application, the mapping between CSI and localization within the cell is clearly a function of the scattering environment, which can not always be assumed constant over long periods of time (due e.g. to new buildings...). Hardware updates, or the aging of the electronic components can also make this mapping vary over time. It is therefore necessary to take into account these slow-scale time variations to properly assess the accuracy of learning-based localization. This calls for updates of the learned function over time, and thus the collection of new data.

If the old training data is not completely stale, it might be useful to re-train the neural network using the merged old and new training datasets. In order to reduce the complexity of this re-training, update mechanisms (such as \cite{Liang06_sequential_learning} for ELMs) can readily be used to provide an online solution.
%(Luis' suggestion): 

The question of when to trigger an update (through the collection of new training data) can be answered by using the theoretical framework of \cite{Louart_etal_RMT_neural_networks_AP17}, e.g., by checking how good the new training data can be predicted by the old one.
In the case of crowdsourcing, the system can include on-demand acquisition of training data through a mechanism whereby data collection is enabled if the measured CSI for a crowdsourcing user can not be accurately matched to its location.

\subsection{Channel frequency selectivity}
\noindent Modern cellular networks rely on some form of frequency-division multiple access (FDMA) to multiplex the users in a cell. As a consequence, for a given user, the channel estimate is available only for a fraction of the spectrum (typically, a group of OFDM subcarriers, or \emph{resource block group}). The consequence is that the channel covariance matrix associated to a given user can only be estimated for a fraction of the spectrum. If the total system bandwidth is large, this covariance matrix can not be assumed constant in the frequency domain. This complicates the learning process, since the mapping between channel covariance and UE location will depend on the considered frequency. One approach is to fractionate the system bandwidth into subbands in which the channel covariance can be assumed constant, and repeat the learning process independently for each subband. Another solution is to add a pre-processing step consisting in transposing (in the frequency domain) the measured covariance matrix to a reference frequency (see e.g. \cite{YantaoHan_FDD_covariance_reciprocity_CHINACOM10,FDD_covariance_interpolation_Globecom2015,Haghighatshoar_covariance_interpolation_2018,Miretti_etal_FDD_covariance_conversion_ICASSP18}), and to let the neural network operate solely on the reference frequency.

\subsection{Channel covariance estimation}
\noindent A full estimate of the uplink channel matrix is not always available at the BS; in particular this is the case for multiple-antenna UEs when precoded reference signals are in use---in this case, only the product between the precoder and the instantaneous channel can be estimated. Another case is when channel estimation is performed for multiple users simultaneously using non-orthogonal reference symbols \cite{Efficient_CSI_acquisition_WSA2017}. However, even in these cases where the complete instantaneous CSI matrix is not estimated, it is generally possible to obtain unbiased estimates of the covariance matrix of each user, at the cost of more complex processing \cite{covariance_estimation_EUSIPCO17}. \\

%\section{Open Questions}
%\label{sec_questions}

%About the approach and the performance limits:
%\begin{itemize}
%\item what is the influence of the considered antenna number on accuracy?
%\item narrowband vs. wideband data performance?
%\item Due to scheduling, users can be active only in a subband in the LF band, not always the same. Is covariance consistent across subbands? If not, can we transpose the covariance matrix to a reference subband? \cite{YantaoHan_FDD_covariance_reciprocity_CHINACOM10} .
%\item data ``aging:'' for how long is the training valid?
%\item 3D positioning
%\end{itemize}

%\begin{appendices}
%\end{appendices}
%\balance

% \newpage
\bibliographystyle{IEEEtran}

%\tiny
\bibliography{refs}
\end{document}